\documentclass[11pt]{article}
\usepackage{moriond,epsfig}

\def\Journal#1#2#3#4{{#1} {\bf #2}, #3 (#4)}

\def\NIM{\em Nucl. Instrum. Methods}

\def\PRL{\em Phys. Rev. Lett.}
\def\PRD{{\em Phys. Rev.} D}

\def\be{\begin{equation}}
\def\ee{\end{equation}}
\def\bea{\begin{eqnarray}}
\def\eea{\end{eqnarray}}

\begin{document}
\hfill{\bf BRIS/HEP/2001-01}\\
\vspace*{4cm}
\title{ELECTROWEAK RESULTS FROM SLD}

\author{ N. de Groot}

\address{H.H. Wills Physics Laboratory, Bristol University,\\
Tyndall Avenue,\\ Bristol BS8 1TL, U.K.}

\maketitle\abstracts{
In this paper we present 
three updates to heavy flavour results from the SLD
detector at SLAC. These results are preliminary, based on
our full 1993-1998 dataset of 550 000 hadronic $Z^0$ decays  produced
with an average electron polarisation of 73\%. The new measurements
are of the hadronic branching fractions into heavy quarks ($R_b, R_c$),
the $b$ quark asymmetry ($A_b$) using jet charge, and the heavy quark
asymmetries ($A_b$ and $A_c$) using  vertex charge and kaons.
}

\section{Introduction}
Due to the unique features the SLC/SLD complex, the SLD experiment is able
to contribute several state-of-the-art electroweak and b-physics
measurements.   These well-known features are:
(a) a high (75\%), precisely measured (${{\delta \cal P} \over{P}} \sim 0.5\%$) 
longitudinal e- polarisation,
(b) a small and stable $e^+e^-$ luminous region (1.5 by 0.8 by 700 $\mu m$)
and a uniquely precise CCD-based vertex detector 
and (c) Good particle identification with a Cerenkov Ring Imaging Device (CRID).
In this paper we will summarise preliminary electroweak results obtained
with the SLD detector~\cite{sld}. In all analyses we are using our standard
hadronic event selection.

\section{Partial Width $R_b$ and $R_c$}
\label{sec:rb}

  An observable providing strong constraints to the Standard
Model is the partial width of the $Z^0$ into $b\:\bar{b}$ final states
$R_b = \Gamma(Z^0\rightarrow b\bar{b})/\Gamma(Z^0\rightarrow \mbox{hadrons})$.
The measurement proceeds by first selecting a pure sample of
$Z^0\rightarrow \mbox{hadrons}$ events. The events are divided in two 
hemispheres, and secondary (and tertiary) vertices are found in the
hemispheres using a topological vertex algorithm~\cite{zvtop}. This
algorithm finds vertices in 75\% of the $b$ hemispheres and in 30\% of
the charm hemispheres. The next step is to calculate the invariant mass of
all tracks in the vertex, assuming they are pions, and correcting for 
 the missing transverse momentum. This quantity is called the vertex mass
and is the starting point for inclusive $b$ and $c$ quark tagging. Those
hemisphere with a secondary vertex are passed through a feed-forward 
neural network. The input variables are the corrected vertex mass, 
vertex momentum, decay length and number of tracks in the vertex, 
the output is 0.05 for charm
events and 0.95 for $b$'s. For $uds$ a target value of 0.5 is used.

\begin{figure}[htb]
\begin{center}
\epsfig{figure=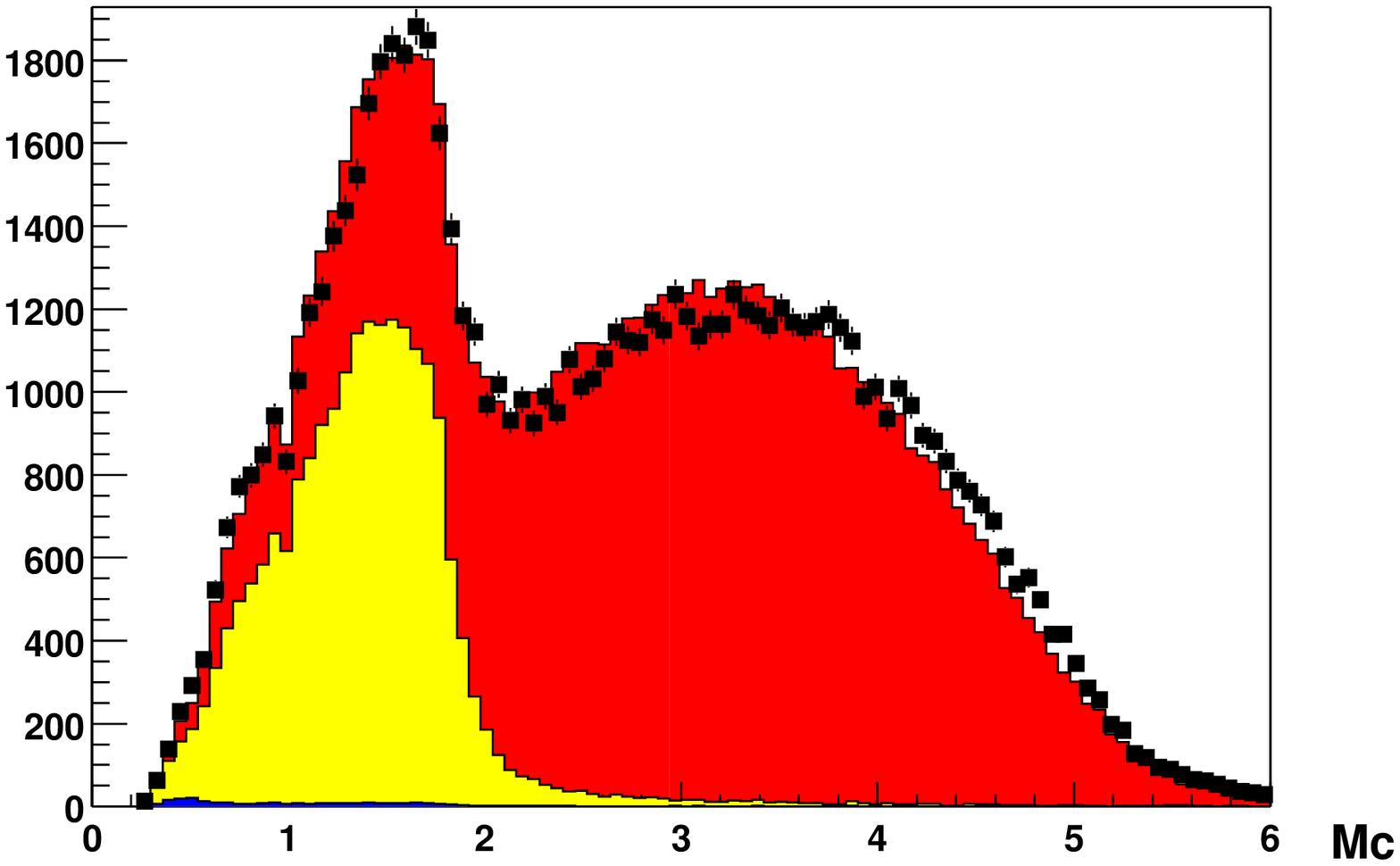,height=2.0in}
\epsfig{figure=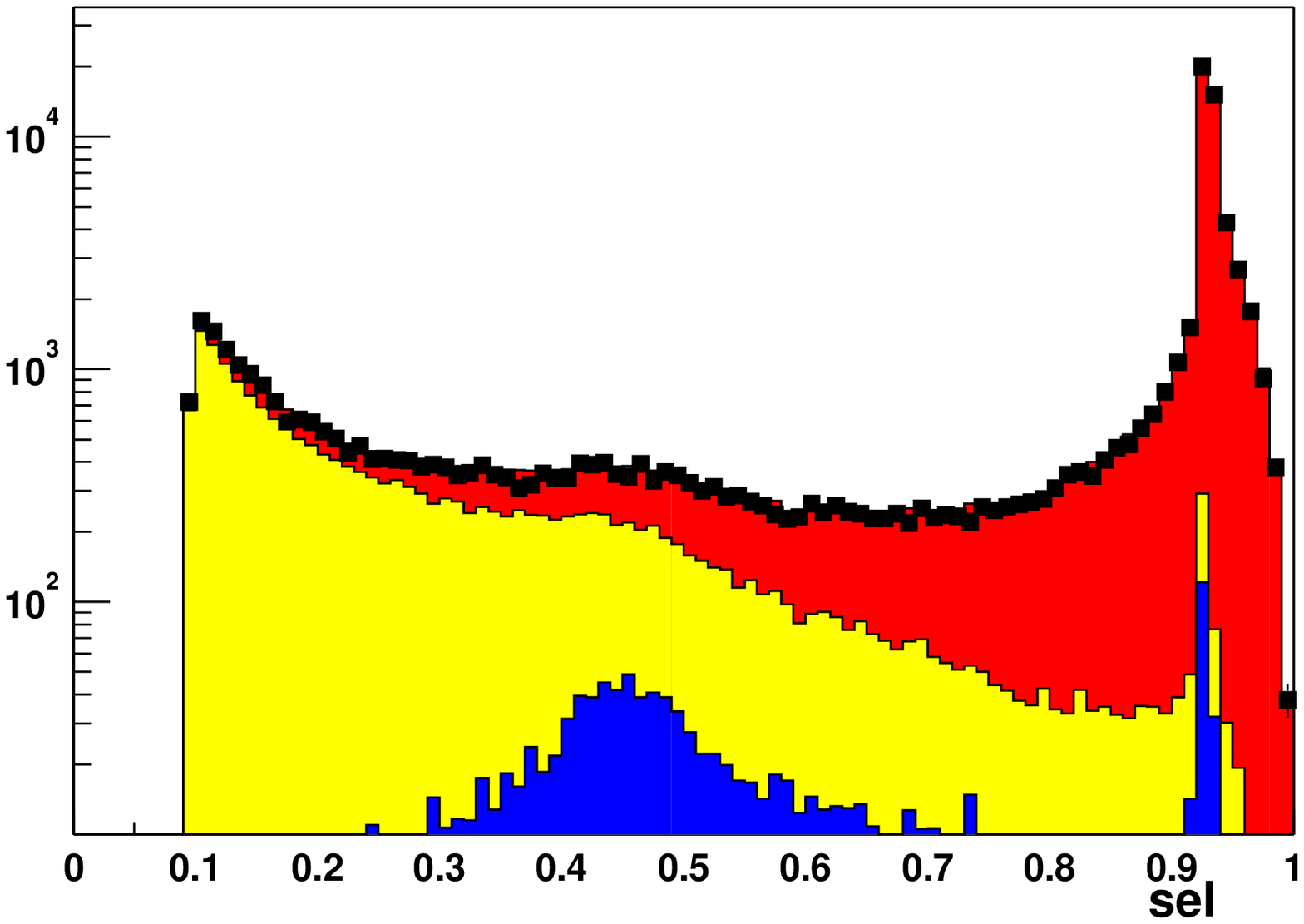,height=2.0in}
\caption{Distribution of the vertex mass (left) and 
the neural network (right) output for data and Monte Carlo.
$b$ events are dark gray, $c$ light, $uds$ black.
\label{fig:nnsel}}
\end{center}
\end{figure}

Then each event hemisphere is $b$-tagged independently for the presence of
a $B$ hadron decay. To be tagged as a $b$, vertices are required to have
a neural network output, $sel$, greater than 0.75.

By computing both the rate for tagging hemispheres 
and the rate for tagging both hemispheres in an event,
one can extract both the value of $R_b$ and the efficiency
of the hemisphere tag $\epsilon_b$~\cite{rb}.
The Monte Carlo is used as input for the charm and $uds$ efficiencies
so it is very important to increase the purity of the $b$-tag to
reduce the systematic uncertainties due to modelling of the charm production
and decay in the Monte Carlo. Our $R_b$ results, combining this new measurement on the 96-98 data with a previously published one on the 93-95 
data~\cite{rb} is:
\[
  R_b = 0.21705 \pm 0.00094 {\mbox{(stat)}} \pm 0.00079 {\mbox{(sys)}}
\]
The corresponding $b$-tag purity
is estimated to be 98.3\%  with an efficiency of 61.7
The dominant systematic errors are coming from uncertainties in the
event selection bias and the gluon splitting into heavy quark pairs.

The measurement of $R_c$ is more complicated. Here we use 
a multi tag method. Next to the b-tag (tag 4), we define a charm tag (1)
for $ 0 < sel < 0.3 $ and a charm-like tag (2) with $ 0.3 < sel < 0.5$ and
a b-like tag (3) with $ 0.5 < sel < 0.75$. With tag~0 being the no tag
category we can write the double and mixed tag fractions as:

\begin{eqnarray}
G_{ii} &=& (\epsilon_b^i)^2 \lambda_b^{ii} R_b
    + (\epsilon_c^i)^2  \lambda_c^{ii} R_c 
    + (\epsilon_{uds}^i)^2 \lambda_{uds}^{ii} (1-R_b-R_c)
  \\ \nonumber
G_{ij} &=& 2 [ \epsilon_b^i \epsilon_b^j \lambda_b^{ij} R_b
    + \epsilon_c^i  \epsilon_c^j \lambda_c^{ij} R_c 
    + \epsilon_{uds}^i \epsilon_{uds}^j \lambda_{uds}^{ij} (1-R_b-R_c) ]
  \label{Equ_Rcm} 
\end{eqnarray}

We take the light quarks efficiencies and the correlations $\lambda$ between 
the hemispheres from Monte Carlo.
Solving Eqs.(\ref{Equ_Rcm}) with a $\chi^2$ fit yields the preliminary result:
\[
  R_c = 0.1757 \pm 0.0032 {\mbox{(stat)}} \pm 0.0024 {\mbox{(sys)}}
\]
The charm tag efficiency is 17.9~\%, measured
from data and a charm purity is 84.5~\%.
The systematic error is dominated by the uncertainty on the 
$uds$ background and the gluon splitting into $c\bar c$ pairs.

\section{Heavy Quark Asymmetries}

Three different techniques have been used to measure the $b$-quark
asymmetry parameter $A_b$. These differ mostly by the method used to
determine which hemisphere contains the primary $b$ quark:
jet charge~\cite{abjc,ab2}, kaon charge, or vertex charge~\cite{abvc}.
The first step (after hadronic event selection) in each analysis is
to tag $Z^0 \rightarrow b \bar{b}$ events either by 
by applying a topological vertex mass cut of a neural network cut.

$A_b$ is determined with a fit to the $\cos\theta$-dependent
differential cross section with the electron
beam polarisation as input. The thrust axis direction is used to
provide the primary quark axis. 

\subsection{$A_b$ with Jet Charge}
\label{Sec_AbJetC}

The $b$/$\bar{b}$ tag is provided by a momentum-weighted track charge
defined as
\begin{equation}
  Q_{\mbox{diff}} = Q_b - Q_{\bar{b}} =
  -\sum Q_{\mbox{track}} \left|\vec{p} \cdot \hat{T}\right|^\kappa
  \mbox{sign}\left(\vec{p} \cdot \hat{T} \right) ,
\end{equation}
where $\vec{p}$ is the 3-momentum of each track and $Q_{\mbox{track}}$ its
charge, $\hat{T}$ is the direction of the event thrust axis, and
the coefficient $\kappa = 0.5$ is chosen such as to maximise the
analysing power of the tag. The quantity $Q_{\mbox{diff}}$
represents the difference between the momentum-weighted charges in the
two hemispheres.
A related quantity -- the sum of the momentum-weighted charges in the
two hemispheres -- is defined as follows
\begin{equation}
  Q_{\mbox{sum}} = Q_b + Q_{\bar{b}} =
  \sum Q_{\mbox{track}} \left|\vec{p} \cdot \hat{T}\right|^\kappa .
\end{equation}

Due to the nearly 100\% efficiency, it is possible and advantageous
to perform a self-calibration technique, i.e. the analysing power
of the tag can be determined directly from the data.
The self-calibration is obtained by comparing the charges $Q$ in the
two hemispheres or more specifically by comparing the widths of the
$Q_{\mbox{diff}}$ and $Q_{\mbox{sum}}$ distributions.

A likelihood analysis yields
\begin{equation}
  A_b = 0.906 \pm 0.022 \mbox{(stat)} \pm 0.023 \mbox{(sys)},
\end{equation}
where the main contribution to the systematic error comes
from the statistical uncertainty in determining
the value of the analysing power of the tag from the data.

\subsection{$A_b$ and $A_c$ with vertex charge and kaons}
A sample of $b$ and charm events is selected using the neural network
tag. We require $sel > 0.9$ for a $b$-tag and $sel < 0.4$ and $P_{vtx} >5$
for a charm tag.
For those events that have a charged vertex, the sign of the charge is
used to tag the charge of the charm quark. For the ones with one or more
charged kaons attached to the vertex we use the charge of the kaon.
In the case these two disagree, we discard the hemisphere. 
We take $R_b$, $R_c$, the $uds$ efficiencies and the hemisphere correlations
from Monte Carlo and fit for the $b$ and $c$ efficiencies for right and
wrong sing tags.
We extract $A_b$ and $A_c$ from an unbinned likelihood analysis
to the full differential cross section.
\begin{equation}
  A_b = 0.913 \pm 0.019 \mbox{(stat)} \pm 0.018 \mbox{(sys)},
\end{equation}
\begin{equation}
  A_c = 0.674 \pm 0.029 \mbox{(stat)} \pm 0.025 \mbox{(sys)}.
\end{equation}
The systematic errors are dominated by self calibration statistics.

The above three $A_b$ analyses are essentially uncorrelated and can
be combined to yield an SLD average of
$A_b(\mbox{SLD}) = 0.912 \pm 0.021$.
Similarly, the lepton $A_c$ measurement can be combined with a
previous measurement based on semi-exclusive reconstruction of
$D^+$ and $D^*+$ mesons to yield an SLD average of
$A_c(\mbox{SLD}) = 0.671 \pm 0.027$.

\section{Conclusions}

We have presented updates of the measurements of $R_b$, $R_c$, $A_b$ with
jet charge and vertex charge and $A_c$ with vertex charge. $R_c$ is the
most precise measurement in the world, and $R_b$ is second in precision.
The asymmetry measurements are unique, thanks to the polarised beam of the SLC.
Other SLD measurements can be found in table~1.

\begin{table}[t]
\caption{Summary of the SLD electroweak measurements
\label{tab:sum}}
\vspace{0.4cm}
\begin{center}
\begin{tabular}{|c|c|l|}
\hline
{ Observable } & {Prelim. Result} & {comments}\\
\hline
{ $sin^2\theta_W $} & {$0.23092\pm0.00027$} & {SLD average~\cite{al,al2}}\\
\hline
{ $A_c$ } & {$0.672\pm0.029\pm0.023$} & vertex/K charge~\cite{abvc}\\
{ $A_c$ } & {$0.685\pm0.052\pm0.038$} & inclusive $D^*$~\cite{acd}\\
{ $A_c$ } & {$0.690\pm0.042\pm0.022$} & exclusive $D$~\cite{acd}\\
{ $A_c$ } & {$0.634\pm0.051\pm0.064$} & leptons~\cite{abl}\\
{\bf $A_c$ } & {$0.671\pm0.027$} & SLD Average\\
\hline
{ $A_b$ } & {$0.906\pm0.022\pm 0.023$} & jet charge~\cite{abjc}\\
{ $A_b$ } & {$0.911\pm0.019\pm 0.018$} & vertex/K charge~\cite{abvc}\\
{ $A_b$ } & {$0.926\pm0.030\pm 0.024$} & leptons~\cite{abl}\\
{ \bf $A_b$ } & {$0.912\pm0.021$} & SLD Average\\
\hline
{ $A_s$ } & {$0.895\pm0.066 \pm 0.063$}&~\cite{as}\\
\hline
{ $R_c$ } & {$0.1757\pm0.0032\pm0.0024$} &
{These observable are}\\
{ $R_b$ } & {$0.21705\pm0.00094\pm0.00079$} &
 {consistent with the SM.}\\
\hline
\end{tabular}
\end{center}
\end{table}

\section*{Acknowledgements}
I would like to thank the SLC and SLD people for the successful operation.
Thanks to Su Dong, Victor Serbo, Sean Walston and Thomas Wright for
their input to this manuscript. This work was supported in part by DOE 
Contract DE-AC03-76SF00515(SLAC).


\section*{References}
\begin{thebibliography}{99}
\bibitem{sld} K. Abe {\it et al}, \Journal{\PRD}{53}{1023}{1996}.
\bibitem{zvtop} D. Jackson, \Journal{\NIM}{A338}{247}{1997}.
\bibitem{rb} K. Abe {\it et al}, \Journal{\PRL}{80}{660}{1998}.
\bibitem{al} K. Abe {\it et al}, \Journal{\PRL}{84}{5945}{2000}.
\bibitem{al2} K. Abe {\it et al}, \Journal{\PRL}{86}{1162}{2001}.
\bibitem{abvc} K. Abe {\it et al}, SLAC-PUB-8542, Contributed to ICHEP 2000.
\bibitem{abjc} K. Abe {\it et al}, SLAC-PUB-8644, Contributed to ICHEP 2000.
\bibitem{ab2} K. Abe {\it et al}, \Journal{\PRL}{81}{942}{1998}.
\bibitem{abl} K. Abe {\it et al}, \Journal{\PRL}{83}{3384}{1999}.
\bibitem{as} K. Abe {\it et al}, \Journal{\PRL}{85}{5059}{2000}.
\bibitem{acd} K. Abe {\it et al}, \Journal{\PRD}{63}{32005}{2001}.

\end{thebibliography}
\end{document}